\documentclass[useAMS,usenatbib]{mn2e}
\usepackage{graphics}
\usepackage{rotating}
\usepackage{epsfig}
\usepackage{color}
\usepackage{natbib}
\def\be{\begin{equation}}
\def\ee{\end{equation}}
\def\bea{\begin{eqnarray}}
\def\eea{\end{eqnarray}}

\title[]{GeV Emission from neutron-rich internal shocks of some long Gamma-ray Bursts}
\author[]{Rong-Rong Xue$^{1,2}$\thanks{E-mail: rrxue@pmo.ac.cn}, Yi-Zhong Fan$^{3,1}$\thanks{E-mail: yizhong@nbi.dk} and Da-Ming
Wei$^{1,4}$\thanks{E-mail: dmwei@pmo.ac.cn}\\
$^1$ Purple Mountain Observatory, Chinese Academy of Sciences,
Nanjing 210008, China.\\
$^2$ Graduate School, Chinese Academy of Sciences,
Beijing 100039, China.\\
$^3$ Niels Bohr Institute, Niels Bohr International Academy, Blegdamsvej 17, DK-2100 Copenhagen, Denmark.\\
$^4$ Joint Center for Particle Nuclear Physics and Cosmology of
Purple Mountain Observatory - Nanjing University, Nanjing 210008,
China.}
\date{Accepted ......
Received ......; in original form ......}

\begin{document}

\maketitle
\begin{abstract}
In the neutron-rich internal shocks model for gamma-ray bursts
(GRBs), the Lorentz factors (LFs) of ion shells are variable, and so
are the LFs of accompanying neutron shells. For slow neutron shells
with a typical LF of approximate tens, the typical $\beta$-decay
radius is $\sim 10^{14}-10^{15}$ cm. As GRBs last long enough
[$T_{90}>14(1+z)$ s], one earlier but slower ejected neutron shell
will be swept successively by later ejected ion shells in the range
$\sim10^{13}-10^{15}$ cm, where slow neutrons have decayed
significantly. Part of the thermal energy released in the
interaction  will be given to the electrons. These accelerated
electrons will be mainly cooled by the prompt soft $\gamma-$rays and
give rise to GeV emission. This kind of GeV emission is particularly
important for some very long GRBs and is detectable for the upcoming
satellite {\it Gamma-Ray Large Area Space Telescope} (GLAST).
\end{abstract}

\begin{keywords}
Gamma Rays: bursts--ISM: jets and outflows--radiation mechanism:
nonthermal
\end{keywords}

\section{introduction}
As realized firstly by Derishev, Kocharovsky $\&$ Kocharovsky
(1999a), the fireball of Gamma-ray Bursts (GRBs) may contain a
significant neutron component in several progenitor scenarios. For
the neutron-rich fireball, the acceleration will be modified
(Bachall \& M\'{e}sz\'{a}ros 2000, hereafter BM00) and there could
be many novel observational signatures: (1) If the neutron abundance
is comparable to that of proton, the inelastic collision between
differentially streaming protons and neutrons in the fireball will
provide us observable 5-10 GeV neutrinos/photons (Derishev et al.
1999a; BM00; M\'{e}sz\'{a}ros \& Rees 2000; Koers \& Giannios 2007);
(2) The very early afterglow emission of GRBs would be modified
significantly because of the $\beta-$decay ($n\rightarrow
p+e^{-}+\nu_{\rm e}$) of the ``fast" neutrons carried in the outflow
(Derishev et al. 1999b; Pruet $\&$ Dalal 2002; Beloborodov 2003b;
Fan, Zhang \& Wei 2005b); (3) In the neutron-rich internal shock
model, for GRBs with a duration $>20(1+z)$ s, the $\beta-$decay
products of the earlier ``slow" neutrons with a typical Lorentz
factor (LF) $\sim $ tens will interact with the ion shell ejected at
later times and give rise to detectable UV/optical flashes (Fan \&
Wei 2004; Fan, Zhang \& Wei 2005a); (4)  As shown in Vlahakis et al.
(2003), for a magnetized neutron-rich outflow, the neutrons can
decouple from the protons at a LF $\Gamma$ in the tens, while
protons, which are collimated to a narrower angle by the
electromagnetic force, continue to be accelerated to a value of
$\Gamma$ in the hundreds. As a result, a two-component jet might be
formed (Vlahakis et al. 2003; Peng et al. 2005)\footnote{ In this
case, the protons and neutrons are separated in the narrow and wide
jet components, respectively. The decay products of the neutrons
have no interaction with the protons collimated in the narrow core.
If this is true, the discussion in this work is invalid.}. Jin et
al. (2007) claimed that the X-ray flat segment of short GRB 051221
was attributed to such a scenario. (5) Razzaque \& M\'esz\'aros
(2006) calculated the synchrotron radiation of the electrons
resulting in the $\beta-$decay and found unique temporal and
spectral signature.

In this work, we show that in the neutron-rich (unmagnetized)
internal shock model, besides the possible UV/optical flashes
predicted in Fan \& Wei (2004), there could be strong GeV
emission. This high energy radiation component is a result of the
inverse Compton cooling of the electrons due to the spacetime
overlapping between the prompt soft $\gamma$-ray photon flow and
the interaction region between the proton shell and decay products
of the neutron shell. Such a cooling process, suppressing the
UV/optical emission significantly, has been taken into account by
Fan et al. (2005a). These authors, however, did not calculate the
possible GeV emission.

\section{GeV Emission from neutron-rich internal shocks}
\subsection{Neutron-rich internal shocks}
In the standard baryonic fireball model, the prompt emission of long
GRBs are likely to be powered by the violent collision of shells
with different LFs (Paczy\'nski \& Xu 1994; Rees \& M\'esz\'aros
1994 ). Until now, the practical  LF distribution of shells in
neutron-free internal shocks model is unknown, let alone the case of
neutron-rich. To reach a high efficiency of radiation, significant
difference of LFs  between two shells with approximately the same
mass ($M$) are favored (Piran 1999).

The unmagnetized outflow is controlled by the dimensionless entropy
$\eta=L/(\dot{M}c^2)$ at $r_0$, where $L$ being the total luminosity
of the ejecta and $r_0$ being the radius of the central engine. The
decoupling of protons and neutrons in a fireball occurs in the
coasting or accelerating regime, depending on whether $\eta$ is
below or above the critical value $\eta_\pi\simeq 2.2\times
10^2L_{51} ^{1/4}r_{0,7}^{-1/4}[(1+\xi)/2]^{-1/4}$, $\xi$ being the
ratio of number density of neutrons to protons (BM00). The
convention $Q_{ x}=Q/10^{ x}$ has been adopted in this Letter, in
units of cgs. If a slow shell is concerned with $\eta_s<\eta_{\pi}$,
neutrons are coupled with protons until the end of the acceleration
episode. However, for a fast shell \footnote{In some afterglow
modeling, initial Lorentz factors of the fireballs $\sim
\sqrt{\eta_f \eta_s} \sim 200-400$ were derived (e.g., Zhang et al.
2006; Molinari et al. 2007). For $\eta_s \sim {\rm tens}$, we have
$\eta_f \sim 1000$. So the assumption $\eta_s <\eta_\pi <\eta_f$ is
reasonable.}, $\eta_f>\eta_{\pi}$, neutrons decouple from protons,
keeping the velocity at that time, while protons can be accelerated
to higher velocity. As a result, $\Gamma_{n,s}=\Gamma_{p,s}=\eta_s$
for slow shells and $\Gamma_{n,f}<\Gamma_{p,f}$ for fast shells,
where the subscripts $p$, $n$ represent the proton /neutron
component; $f,~s$ represent the fast/slow shells respectively
(BM00). So far we have four kinds of shells, i.e., slow neutron/ion
shells and fast neutron/ion shells.

Similar to the standard neutron-free fireball, the fast ion shell
will catch up with the slower but earlier one and forms a new one
moving with a LF $\Gamma_{m}\sim {\rm a~few~hundred}$ at a radius
$R_{\rm int}\sim 10^{13}{\rm cm}$,  suppose that the ejection
time-lag between these two shells is about 0.01s (hereafter the
formed ``new'' shell is named as the ``i-shell''). The $\beta$-decay
radius of a neutron shell (hereafter, the ''n-shell") reads as
\begin{equation}
R_{ \beta}\approx 2.6\times 10^{13}\Gamma_{ n}~{\rm cm}.
\end{equation}
For $\Gamma_{n}\sim {\rm tens-hundreds}$, $R_{ \beta}\gg R_{\rm
int}$, so the $\beta-$decay of these fast neutrons won't influence
the physical process taking place at $R_{\rm int}$ significantly.

However, the much later ejected i-shell will catch up the earlier
slow n-shell at a radius $R_{\rm cat}\approx 2\Gamma_{n,s}^2c \delta
T/(1+z)$, where $\delta T$ being the ejection time-lag between the
earlier slow n-shell and the later i-shell and $z$ being the
redshift of GRB. As long as $R_{\rm cat}$ is comparable with $R_{
\beta,s}$, i.e.,
\begin{equation}
\delta T\geq 14(1+z){(\Gamma_{ n,s}/30)^{-1}}~{\rm sec},
\end{equation}
the decay-product of earlier ejected but slower neutron shells will
be swept orderly by the later ejected ion shells in a range $\sim
R_{\rm cat}-R_{ \beta,s}$ and interesting UV/optical and GeV
emission are produced.

In this Letter, following Fan \& Wei (2004), we assume: (1) The LF
of shells has bimodal distribution, $\rm \Gamma_{ej}=\eta_f$ or $\rm
\Gamma_{ej}=\eta_s$ with equal probability, which is favored by its
relatively high efficiency of energy conversion and high peak energy
of the internal shock emission (Guetta, Spada \& Waxman 2001). (2)
$\xi=1$. As a result, the ratios between the mass of the fast
neutrons, slow neutrons and i-shells are $\sim 1:1:2$. (3) Proton
and neutron shells move with LF $
\Gamma_{p,f}\sim1000,\Gamma_{n,f}\sim 200$ for fast shells and $
\Gamma_{p,s}=\Gamma_{n,s}\sim30$ for slow shells. After the merger
of a pair of fast/slow ion shells, the resulting i-shell moves with
\cite{Piran99}
\[
\Gamma_{\rm i} \approx
\sqrt{\frac{M_{p,f}\Gamma_{p,f}+M_{p,s}\Gamma_{p,s}
}{M_{p,f}/\Gamma_{p,f}+M_{p,s}/\Gamma_{p,s}}} \approx 200.
\]
(4) The ejecta expands into low density interstellar medium (ISM),
as mostly found in afterglow modeling (Panaitescu $\&$ Kumar 2002).
The ISM has been swept by the early and fast ions as well as the
decayed products of fast neutrons, so the decay products of slow
neutrons move freely.

Notice that in front of the i-shells ejected at a time $\geq
14(1+z){(\Gamma_{ n,s}/30)^{-1}}$ sec, there are hundreds of
decaying n-shells. With assumptions made before, the whole process
of each i-shell interacting with these decaying n-shells is rather
similar. For convenience, in our following treatment, the discrete
interaction of each i-shell with these decaying n-shells has been
simplified as the situation of an i-shell sweeping a moving proton
trail (the LF of which is $\Gamma_{ n,s}$) with a number density
(Fan \& Wei 2004)
\begin{equation}
n\approx {\Gamma_{n,s}M_{n,s}\over 2\pi R^2m_{ p}R_{
\beta,s}}\label{Eq-n}
\end{equation} continually, where $m_{ p}$ is the rest mass of protons
and $M_{n,s}\propto \exp(-R/R_{ \beta,s})$ is the mass of the
decaying/slow neutron shell.

\subsection{dynamics}
The dynamical evolution of the i-shell is governed by the energy
conservation of the system and can be estimated as (Fan \& Wei
2004):
 \begin{equation}
 \frac{d\gamma}{dm}=-\frac{\gamma\gamma_{\rm rel}-\Gamma_{n,s}}{M_{\rm i}+m+(1-\epsilon)U}
 \end{equation}
where $\gamma$  represents the Lorentz factor  of the decreasing
i-shell and $\gamma_{\rm rel}\approx (\Gamma_{\rm n,s}
/\gamma+\gamma/\Gamma_{\rm n,s})/2$ indicates the LF of the i-shell
relative to the n-shell. In the adiabatic situation, the radiation
efficiency factor $\it \epsilon$ is assumed as zero. $
U=(\gamma_{\rm rel}-1)m{\rm c^2}$ is the thermal energy in the
comoving frame. The swept mass satisfies $dm=4\pi n m_p \Gamma_{n,s}
R^2 (\beta_{\gamma}-\beta_{\Gamma_{n,s}})dR$. $\beta_{\gamma}$ and
$\beta_{\Gamma_{n,s}}$ are the corresponding velocity of $\gamma$
and $\Gamma_{n,s}$.

\begin{figure}\label{fig:dyn}
\includegraphics[width=250pt]{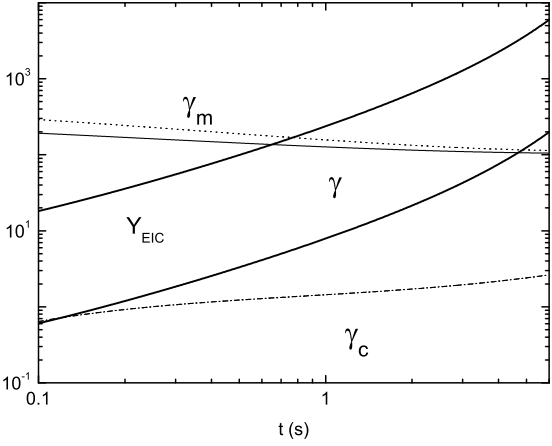}
\caption{The thick solid line represents the evolution of the
external inverse Compton parameter ($ Y_{\rm EIC}$). Here, the
luminosity of the prompt $\gamma$-ray emission is taken as
$\rm10^{51}erg~s^{-1}$. The rapid increase of $Y_{_{\rm EIC}}$ is
due to the exponential decline of the mass of the decay product of
the neutron shell. Note that the upper thick line is the result
assuming a series of parameters as in section 2.2, while the lower
one is for the same parameters except $\epsilon_B\sim
\epsilon_e=0.3$. The solid line indicates the dynamical evolution
of the i-shell ($\gamma$), while the dotted and dashed-dotted line
represent the minimum LF of accelerated electrons ($\gamma_m$) and
the cooling Lorentz factor ($ \gamma_c$), respectively.}
\end{figure}

In the numerical calculation, we take the following parameters: the
mass of i-shell and n-shell $ M_{\rm i}= 2M_{\rm n} \approx
5.6\times10^{27}$g, the initial LF of i-shell $ \Gamma_i \approx
200$, the LF of n-shell $ \Gamma_{n,s}$=30, and the width of n-shell
is taken to be $\sim 3\times10^8$ cm. $\epsilon_e$ and $\epsilon_B$,
the fraction of shock energy  given to the electrons and magnetic
field, are taken as 0.3 and 0.01, respectively. The minimum LF of
the shocked electrons is estimated as
\begin{equation}
\gamma_{ m} \approx \epsilon_e {p-2\over p-1} {m_{ p}\over m_{\rm
e}} (\gamma_{\rm rel}-1)+1,
\end{equation}
where $m_e$ being  the rest mass of electrons and $p$ being the
power-law index of the distribution of accelerated electrons. Our
numerical results are shown in Fig.\ref{fig:dyn}.

The electrons lose their energy via synchrotron radiation and
inverse Compton scattering on the soft $\gamma-$ray photons with a
luminosity $L_\gamma \sim 10^{51}~{\rm erg~s^{-1}}$. The inverse
Compton is usually in the Thompson regime because $\gamma_{ m}
E_{\rm p}/\gamma \sim E_{\rm p}<m_{ e} c^2$, where $E_{\rm p}\sim
200$ keV is the typical peak energy of GRBs (Preece et al. 2000). So
the inverse Compton parameter can be estimated by
\begin{equation}
Y_{_{\rm EIC}}\approx {U_\gamma \over U_B}, \label{parameter}
\end{equation}
where $U_\gamma =L_\gamma /(4\pi R^2 \gamma^2 c)$  and
$U_B=4\epsilon_B \gamma_{\rm rel}(\gamma_{\rm rel}-1)nm_pc^2$. The
numerical estimate of $Y_{\rm _{EIC}}$ has been presented in
Fig.\ref{fig:dyn}. One can see that the cooling of electrons is
dominated by the inverse Compton process. Now the seed photons are
from a region separated from the scattering electrons and along the
direction in which the ejecta moves. Following Fan \& Piran (2008),
we call this kind of inverse Compton scattering as the External
inverse Compton (EIC). The corresponding synchrotron self-Compton
parameter $Y_{_{\rm SSC}}$ can be estimated as follows. Following
Sari \& Esin (2001), in the case of fast cooling (see below for the
reason), we have
\begin{equation}
Y_{_{\rm SSC}}\approx {1\over 1+Y_{_{\rm SSC}}+Y_{_{\rm
EIC}}}{\epsilon_{\rm e} \over \epsilon_{\rm B}},
\end{equation}
which gives (see also Fan \& Piran 2006)
\begin{eqnarray}
Y_{_{\rm SSC}} &\approx& {{-(1+Y_{_{\rm EIC}})+\sqrt{(1+Y_{_{\rm
EIC}})^2+4\epsilon_{\rm e}/\epsilon_{\rm B}}}\over 2}\nonumber\\
&\approx&
\left\{%
\begin{array}{ll}
    {\epsilon_{\rm e}/\epsilon_{\rm B} \over 1+Y_{_{\rm EIC}}}, & \hbox{for $(1+Y_{_{\rm
EIC}})^2\gg 4\epsilon_{\rm e}/\epsilon_{\rm B}$},\\
0.2(1+Y_{_{\rm EIC}}), & \hbox{for $(1+Y_{_{\rm
EIC}})^2\approx 4\epsilon_{\rm e}/\epsilon_{\rm B}$},\\
\sqrt{\epsilon_{\rm e}/\epsilon_{\rm B}}, & \hbox{for $(1+Y_{_{\rm
EIC}})^2\ll 4\epsilon_{\rm e}/\epsilon_{\rm B}$.}
\end{array}%
\right.
\end{eqnarray}
From Fig.\ref{fig:dyn}, {\bf  we see that $(1+Y_{_{\rm EIC}})^2\gg
4\epsilon_{\rm e}/\epsilon_{\rm B}$ } in most cases, so the EIC
emission always dominates the SSC emission.

The cooling Lorentz factor $\gamma_c$ of the electrons can be
estimated as (Piran 1999)
\begin{equation}
 \gamma_c \approx \frac{6\pi m_e c(1+z)}{(1+Y_{_{\rm EIC}})\sigma_T \gamma B^2
 t},
\end{equation}
where $t$ is the observer's timescale, $\sigma_T$ is the Thomson
cross section, and B is the comoving downstream magnetic field $
B\approx\sqrt{32\pi \epsilon_e \gamma_{\rm rel}(\gamma_{\rm rel}-1)n
m_p c^2}$.

As shown in Fig.\ref{fig:dyn}, $\gamma_c\ll \gamma_m$.  The
electrons are in the fast cooling regime.

\subsection{GeV emission}
The electrons accelerated in the interaction between i-shell and the
decay product of n-shells will subsequently Compton scatter the
prompt soft $\gamma-$ray photons, and boost them to an energy
\begin{equation}
{\rm h} \nu_{_{\rm EIC}}\sim \gamma_m^2 E_{\rm p}\sim  2~{\rm GeV}
(\gamma_{ m}/100)^2 (E_{\rm p}/200~{\rm keV}).
\end{equation}

The total energy of the GeV emission can be estimated as follows.
Note that in our simplest model, the energy released in an i-shell
interacting with the $\beta-$decay products of a series of neutron
shells is equivalent to that of the interaction between an ion-shell
and all the decay products of a neutron shell. So the GeV emission
is largely the same as that resulting in the internal shocks at
$R_{\rm \beta,s}$ powered by two ion-shells. Following Piran (1999),
we consider a collision between these two shells with mass $\rm M_i$
and $\rm M_s$ which are moving with different LFs:
$\Gamma_i\gg\Gamma_{n,s}\gg1$. Here, the same parameters are taken
as in section 2.2. The resulting bulk LF $\Gamma_m$ in an elastic
collision is
\begin{equation} \Gamma_m=\sqrt{\frac{M_{\rm i}\Gamma_i+M_{\rm
n}\Gamma_{n,s} }{M_{\rm i}/\Gamma_i+M_{\rm n}/\Gamma_{n,s}}}=108.
\label{eq:gamma}
\end{equation}
The total internal energy (the difference of the kinetic energies
before and after the collision) is
\begin{equation}
E_{\rm the}=\Gamma_i M_{\rm i} {\rm c^2}+\Gamma_{n,s} M_{\rm n} {\rm
c^2}-\Gamma_m (M_{\rm i}+M_{\rm n}){\rm c^2}=3.1\times 10^{50} \rm
erg.
\end{equation}
Then the total energy given to electrons:
\begin{equation}
 E_e=\epsilon_e E_{\rm the}=9.3\times 10^{49}  \rm erg
(\frac{\epsilon_e}{0.3}).
\end{equation}

Since the electrons are in the fast cooling regime, the GeV
radiation efficiency can be estimated as (for $Y_{\rm _{EIC}}\geq
1$)
\begin{equation}
\eta_{_{\rm GeV}}={Y_{\rm _{EIC}} \over 1+Y_{\rm
_{EIC}}}\frac{E_e}{E_{\rm tot}}\approx \frac{E_e}{\Gamma_i M_i {\rm
c^2}+\Gamma_{n,s} M_n {\rm c^2}}.
\end{equation}
With the typical parameters adopted in this work, we get
$\eta_{_{\rm GeV}}\sim 8.5\%$. This is comparable to the typical GRB
efficiency $\eta \sim 20\%$ (Guetta, Spada \& Waxman 2001). We thus
draw the conclusion that  the total energy of our GeV radiation
component is comparable to the total energy of the soft $\gamma-$ray
photons in some very long GRBs.

The EIC photons may be absorbed due to pair production by photons
with energy above $ E_a\sim2(\gamma m_e c^2)^2/(\rm h\nu_{
EIC})\sim 1~ \rm MeV\gamma_2^2/( h\nu_{EIC}/5GeV)$. And the
corresponding optical depth can be estimated as (e.g., Svensson
1987)
\begin{eqnarray} \tau_{\gamma\gamma} &\simeq & \frac{11\sigma_T
N_{>E_a}}{720\pi R^2}\nonumber\\  &\sim &  0.2
R_{15}^{-2}L_{\gamma,51}\delta T (\frac{E_p}{\rm
200keV})^{\beta_{\gamma}-1}\gamma_2^{-2\beta_{\gamma}}(\frac{\rm
h\nu_{EIC}}{\rm 5 GeV})^{\beta_{\gamma}}\nonumber
\end{eqnarray}
where
$N_{>E_a}=\frac{\beta_{\gamma}-1}{\beta_{\gamma}}(\frac{E_p}{E_a})^{\beta_{\gamma}}\frac{L_{\gamma}\delta
T}{E_p}$ is the total number of photons of the prompt emission
satisfying ${\rm h}\nu>E_a$, $\delta T \sim R_{\beta,s}/
(2\gamma^2 c)$ is the timescale of the late prompt emission
overlapping with EIC emission and its power-law index
$\beta_{\gamma}\sim 1.2$ has been used to get the analytical
coefficient. The ultimate optical depth is $\sim 0.2$ and the
spectrum  around 5 GeV would not suffer a significant absorption.

\begin{figure}\label{fig:spectrum}
\begin{center}
\includegraphics[width=250pt]{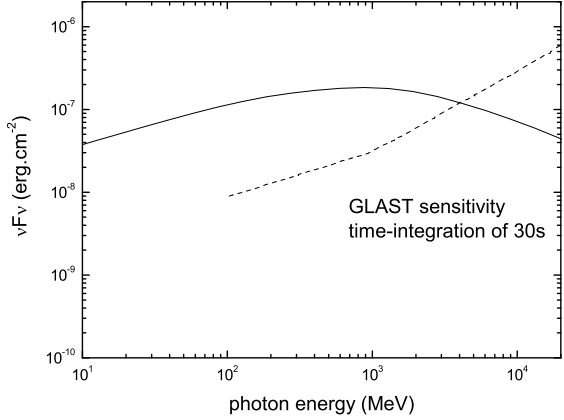}
\end{center}
\caption{The solid line represents the time-integrated GeV
emission spectrum resulting from prompt soft $\gamma$-ray emission
upscattered by electrons in the neutron-rich internal shocks
model. GLAST sensitivity with corresponding exposure time is also
labelled as the dashed line (Galli \& Guetta 2008). }
\end{figure}

Now we estimate the detectability of this GeV component by the
upcoming satellite GLAST. The Large Area Telescope on board covers
the energy range from 20 MeV to 300 GeV. The effective area around 1
GeV is $S_{\rm eff} \sim \rm 10^4 cm^2$. So the expected number of
the GeV photons is
\begin{eqnarray}  N_{\rm det} &\approx & {\eta_{\rm _{GeV}} \over \eta} \frac{E_\gamma}{4\pi
D_L^2 {\rm h}\nu_{_{\rm EIC}}}S_{\rm eff} \nonumber\\
&\sim & 5 {E_\gamma \over 10^{52}~{\rm erg}}({\rm h\nu_{_{
EIC}}\over 1~{\rm GeV}})^{-1}.
\end{eqnarray}
Usually at least 5 high energy photos are needed to claim a
detection (e.g., Zhang \& M\'eszaros 2001), so we conclude that
this component is detectable for some bright long GRBs. Please see
Fig.\ref{fig:spectrum} for a numerical example,  which shows $\nu
F_{\nu}\propto \nu^{1/2}$ below the peak as a result of the
scattering electrons in the fast cooling phase.

\section{discussion and conclusion}
The physical composition of GRB outflows is not clear yet. The
outflows, in principle, could be Poynting-flux dominated or
neutron-rich. People found some evidences for either model. For
example, the modeling of optical flashes in some GRBs suggested the
magnetization of the reverse shock regions and favored the
hypothesis that these GRB outflows were at least weakly magnetized,
while the delayed/prompt optical flares detected in GRB 041219a and
some other afterglow modeling got some indication evidences for
neutron-rich outflows (see Zhang 2007 for a review). The physical
composition of GRB outflows may be not universal and more effects
are needed to get some ``unique" signatures of the Poynting-flux
dominated or neutron-rich ejecta. In this work, we calculate the
possible GeV emission signature of the unmagnetized  neutron-rich
internal shocks.
 This high energy component, only emerging in some very long GRBs, is
 accompanying some UV/optical flares (Fan \& Wei 2004) and
is expected to be correlated with the prompt $\gamma$-ray emission
but to have a  $\sim 10(1+z)$ s lag. This is very different from Li
\& Waxman (2008), in which the UV/optical/GeV flares are powered by
a series of weaker and weaker internal shocks at a distance $\sim
10^{15}$ cm and are almost simultaneous with the prompt
$\gamma-$rays without an obvious lag. The time delay of the EIC
emission, caused by the anisotropic emission of the scattered
photons in the rest frame of the shocked medium (see Fan \& Piran
2008 for a recent review), $\sim (1+z)R_{\beta,s}/(2\gamma^2c)$, is
also ignorable in the case we consider.

In some very long bright GRBs, the delayed GeV emission discussed in
this work is energetic enough to be detectable for the upcoming
GLAST satellite (see section 2.3 for details). It, however, might be
outshined by the synchrotron-self Compton (SSC) radiation of the
regular internal shocks at a radius $R_{\rm int}\sim 10^{13}$ cm. So
only for the regular internal shocks having $\epsilon_e \sim
\epsilon_B$, the delayed GeV emission may be the dominant signal. As
shown in Fig.\ref{fig:dyn}, for $\epsilon_e \sim \epsilon_B$,
$Y_{_{\rm EIC}}\gg 1$ at $t\sim R_{\beta,s}(1+z)/2{\gamma}^2 c > 1
s$. So most of the energy generated at a radius $\sim R_{\beta,s}$
is lost into the EIC GeV emission. For Gamma-ray Burst Monitor (GBM;
$<20$ MeV) onboard GLAST, the field of view is all sky not occulted
by the earth and that of the Large-area Telescope (LAT; $30~{\rm
MeV}\sim 300~{\rm GeV}$) is $\sim 2.5$ sr. So $\sim 1/5$ GRBs will
be detected by GBM and LAT simultaneously. With a good quality of
keV$-$10 GeV spectrum, we may be able to check our model.

%%\begin{center}
%\includegraphics[width=250pt]{range.eps}
%\end{center}
%\caption{The dashed line represents the case when $\rm
%\epsilon_e+\epsilon_B=1$ and the solid line indicates the situation
%when the cooling effect from two inverse Compton process are
%comparable. The shaded area is the  parameter range where the EIC
%emission considered in our work dominates the SSC emission .
%}\label{fig:region}
%\end{figure}

\section*{Acknowledgments}
This work is supported by the National Science Foundation (grants
10673034 and 10621303) and National Basic Research Program (973
program 2007CB815404) of China. YZF is supported by a postdoctoral
grant from Danish National Research Foundation and by a special
grant of Chinese Academy of Sciences.

\end{document}